\documentclass[12pt,leqno]{extarticle}
\usepackage[utf8]{inputenc}
\usepackage{geometry}
\usepackage{enumitem,amssymb,amsmath,xcolor,cancel,hyperref,graphicx,etoolbox,setspace,subcaption,lmodern,yfonts,float,colortbl,dsfont,centernot,comment, ulem}
\usepackage[font=small,labelfont=bf,width=0.95\textwidth]{caption}
\usepackage{authblk}
\usepackage{rotating}
\usepackage{indentfirst}

\usepackage{appendix}
\usepackage{makecell}
\usepackage{textcomp}
\usepackage{cite}
\newcolumntype{C}[1]{>{\centering\arraybackslash}m{#1}}
\newcommand{\norm}[1]{\left\lVert#1\right\rVert}

\newcommand{\xx}{\mathbf{x}}
\newcommand{\yy}{\mathbf{y}}

\DeclareMathOperator{\Exp}{\mathbb{E}}
\DeclareMathOperator{\Prob}{\mathbb{P}}
\DeclareMathOperator{\Var}{\mathrm{var}}
\DeclareMathOperator{\Cov}{\mathrm{cov}}

\allowdisplaybreaks
\counterwithin{equation}{section}
\title{Modularity, asymmetry, and polarization shape consensus speed in the voter model}
\author[1]{Madi Yerlanov\footnote{Corresponding author}}
\author[2]{Zachary Kilpatrick}
\author[3]{Nancy Rodr{\'\i}guez}
\affil[1]{University of Colorado Boulder, Department of Applied Mathematics,\authorcr 1111 Engineering Center, Boulder, CO, 80309, \authorcr email: \href{mailto:madi.yerlanov@colorado.edu}{madi.yerlanov@colorado.edu}}
\affil[2]{University of Colorado Boulder, Department of Applied Mathematics,\authorcr 1111 Engineering Center, Boulder, CO, 80309, \authorcr email: \href{mailto:zpkilpat@colorado.edu}{zpkilpat@colorado.edu}}
\affil[3]{University of Colorado Boulder, Department of Applied Mathematics,\authorcr 1111 Engineering Center, Boulder, CO, 80309, \authorcr email: \href{mailto:rodrign@colorado.edu}{rodrign@colorado.edu}}

\date{\today}
\begin{document}
\maketitle
\begin{abstract}
 In populations with community structure, the formation of consensus requires both alignment within and diffusion of beliefs across groups, processes that evolve on distinct time scales. How do modularity, asymmetry, and polarization shape this process?
 We study a variant of the voter model in which a population is divided into two cliques of sizes $N_1$ and $N_2$. At each time step, a pair of nodes is selected; if their binary opinions differ, each agent adopts the opinion of the other with probability $p$. With probability $\alpha$, the pairing occurs with a single clique, and with probability $1-\alpha$, across cliques.
 We analyze how this coupling strength, population imbalance, and initial polarization jointly determine the time to consensus. Formation of consensus generally starts with inter-clique interactions rapidly synchronizing the two cliques’ opinion fractions, after which consensus is reached through a slower diffusion along the synchronized manifold; this slow stage is largely insensitive to $\alpha$ except when the cliques are nearly disconnected.
To analyze these dynamics, we derive stochastic differential equations and Fokker–Planck approximations in the large-population limit, and assess their accuracy against the discrete model. While $\alpha$ primarily affects the fast alignment stage, initially polarized and asymmetric populations exhibit nontrivial effects, including regimes in which an intermediate level coupling minimizes consensus time.
A small-clique scaling analysis reveals that this optimum arises from a competition between fast alignment drift and noise amplification in the smaller group, and provides an approximate decomposition of consensus time into fast and slow contributions.
\end{abstract}

\section{Introduction}
Many systems that display collective decision-making are inherently modular, with interactions occurring more frequently within communities than across them~\cite{girvan2002community,fortunato2010community}.  In social and political opinion dynamics, individuals are often organized into geographical regions or interest-based communities, where internal interactions are common while cross-community interactions occur at lower rates, sometimes facilitated by social media~\cite{centola2018behavior,castellano2009statistical}.  
Indeed, experiments have shown that clustered network structure can accelerate the spread of new behaviors~\cite{centola2010spread}, yet the mechanisms linking modularity to consensus speed remain poorly understood analytically.
Classical voter models, which assume uniform mixing, do not capture how such modularity influences consensus formation or the persistence of disagreement~\cite{liggett2013stochastic,clifford1973model, holley1975ergodic}. Extensions to structured networks have begun to relax these assumptions by incorporating heterogeneous connectivity~\cite{sood2005voter,sood2008voter, antal2006fixation}, community structure \cite{masuda2014voter, gastner2019voter}, and opinion-dependent interaction rules~\cite{deffuant2000mixing, hegselmann2002opinion}, but a general understanding of how weak coupling between subgroups shapes fixation remains incomplete.

Many ecological populations are divided into subgroups that interact more frequently within their own habitat than with others~\cite{grilli2016modularity}. Occasionally, individuals move between habitats or observe and copy behaviors from other groups, thereby enabling the spread of behavioral traits, such as foraging strategies or anti-predator responses, across an entire population~\cite{danchin2004public}.  Over time, one behavior may become {\it fixed}, meaning that nearly all individuals in both groups adopt the same strategy~\cite{bergstrom2002evolution,antal2006fixation}.
The interplay between population structure and fixation dynamics has been studied extensively in evolutionary graph theory~\cite{allen2017evolutionary, hathcock2019fitness}, where timescale separation between local equilibration and global fixation plays a central role~\cite{constable2016demographic, constable2018exploiting}.

At a more abstract level, similar dynamics arise in distributed computing systems, where many independent units must coordinate through local communication~\cite{ghosh2006distributed,olfatisaber2004consensus}. In such systems, nodes often interact in a modular manner, contacting nearby or related nodes more frequently than most nodes in the network. A binary opinion can represent a local decision, protocol choice, or stored value, and pairwise interactions model information exchange between nodes. In this context, consensus corresponds to global agreement across the network, and the strength of coupling between modules plays a critical role in determining how rapidly such agreement is reached~\cite{olshevsky2009convergence}.

In this work, we study a tractable, modular extension of the classic voter model~\cite{redner2019reality,gastner2019voter} to quantify how network modularity impacts consensus formation. We consider two connected communities where inter- and intra-group interactions occur at rates governed by a modularity parameter, individuals hold one of two states, and opinion changes arise through random pairwise interactions with a tunable alignment parameter. Voter models on modular and multi-clique networks have been studied previously, including effects of external sources~\cite{bhat2020polarization}, the interplay between population size and inter-clique connectivity~\cite{gastner2019voter}, and consensus times in structured populations~\cite{masuda2014voter,sood2008voter,antal2006fixation}, with reviews in~\cite{redner2019reality,castellano2009statistical}. Building on this literature, we vary the relative strength of within- and between-community interactions to examine how connectivity and population size shape consensus and fixation times. We analyze these dynamics through agent-based simulations alongside stochastic and continuum approximations~\cite{constable2018exploiting}, systematically relating modular structure to the mechanisms governing consensus.

We begin by introducing a discrete, two-clique voter model and specifying the opinion-update rules that govern interactions within and between communities. From this formulation, we compute the mean and variance of opinion changes and use these quantities to derive a stochastic differential equation approximation in the large-population limit, followed by a fully continuous description in the form of a Fokker–Planck equation.
This continuum formulation provides a natural setting for asymptotic analysis, which we use to uncover the mechanisms underlying consensus formation in large systems. In particular, the analysis reveals a separation of time scales, with rapid alignment between cliques followed by slower diffusion toward global consensus. To assess the validity and limitations of these approximations, we complement the analysis with numerical simulations of the discrete, stochastic, and continuum models, comparing their predictions across a range of parameters. These results highlight how modularity, population imbalance, and interaction strength shape consensus times, and motivate further investigation of data-driven and heterogeneous extensions of voter models.

A key take-home message is that modular structure matters most when the system begins far from symmetry. In near-symmetric settings in which the clique sizes and initial opinion fractions are balanced, the well-mixed model is often the fastest route to consensus, and introducing modularity tends to add an alignment step without improving the slow diffusion to fixation. By contrast, strongly polarized initial conditions expose regimes in which nontrivial modularity can be advantageous: intermediate inter-clique coupling can accelerate the initial equalization of opinion fractions while still allowing efficient drift toward consensus. These effects are especially pronounced in mismatched systems, where a moderately coupled two-module network can enable the larger clique to quickly absorb the smaller one, reducing overall consensus times relative to both extremes of near-disconnection and near-complete mixing. The analysis and simulations that follow make these regimes precise, showing that modularity is neither universally beneficial nor universally detrimental, but instead acts as a context-dependent control parameter whose impact depends critically on initial polarization and population imbalance.

\section{A Modular Voter Model and Mean-Field Limits}
In this section, we introduce a modular extension of the classical voter model and specify the discrete interaction rules governing opinion updates within and between communities. Starting from this stochastic description, we derive diffusion and Fokker–Planck approximations that allow us to analyze consensus formation across different coupling regimes.

\begin{figure}[t!]
    \centering
\includegraphics[width=0.75\linewidth]{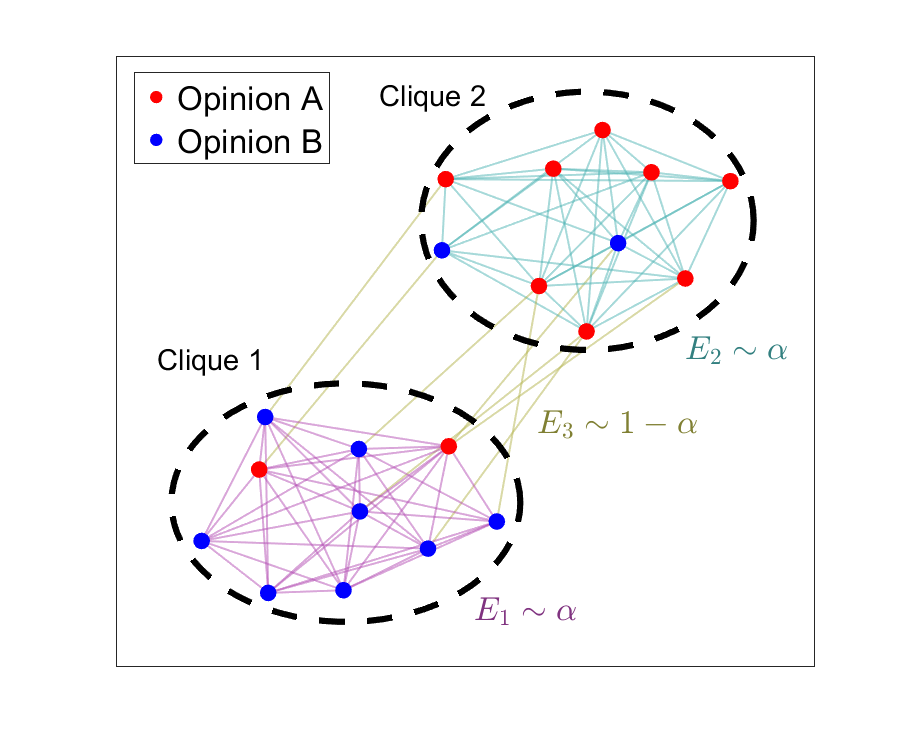}
\vspace{-10mm}
    \caption{{\bf Example of a two-clique network.} Each clique has a fixed number of nodes ($10$ in this case), whose opinions are updated at discrete time steps. Intra- and inter-clique connections were sampled with different weights governed by a modularity parameter, $\alpha=0.9$. We define consensus as a state in which (almost) all nodes have the same color (\textit{i.e.}, almost all red or all blue).}
    \label{fig:networkexample}
\end{figure}

\subsection{Model Definition and Interaction Rules}
We consider a network with $N$ nodes partitioned into two cliques of sizes $N_1$ and $N_2$, with $N_1+N_2=N$. Each node holds a binary opinion, denoted as $-1$ or $+1$ (equivalently, A and B), and we use them interchangeably throughout this paper. Let $x_1:=x_1(t)$ and $x_2:=x_2(t)$ represent the proportions of nodes holding the opinion $-1$ in cliques 1 and 2. We denote the corresponding initial conditions by $y_1:=x_1(0)$ and $y_2:=x_2(0)$.
The dynamics proceed in discrete time steps of duration $\Delta t=2/N$.
At each step, an edge is selected uniformly at random from the network, with the relative frequencies of intra- and inter-clique edges governed by a modularity parameter $\alpha\in[0,1]$. This construction should be understood as a mean-field interaction rule: all node pairs are in principle possible, but intra- and inter-clique pairs are sampled with different weights determined by $\alpha$, rather than through a fixed adjacency structure.  Let $E_1$, $E_2$, and $E_3$ denote the (expected) numbers of edges within clique~1, within clique~2, and between cliques,
\begin{align*}
E_1=\frac{\alpha}{2}N_1(N_1-1),\qquad
E_2=\frac{\alpha}{2}N_2(N_2-1),\qquad
E_3=(1-\alpha)N_1N_2.
\end{align*}
Let $E=E_1+E_2+E_3$, and define $\gamma_i=E_i/E$, the probability that a randomly selected edge is of type $i$.
If the selected edge connects two nodes holding opposite opinions, each node independently switches its opinion with probability $p$; if both nodes hold the same opinion, no update occurs.
We refer to this parameter as \textit{flip/switch probability}. Pair selection is governed by the \textit{modularity parameter} $\alpha$, which controls the relative frequency of within- versus between-clique interactions.
When $N_1=N_2$, choosing $\alpha=\tfrac12$ yields uniform sampling over all unordered node pairs and recovers the standard well-mixed (single-clique $N_1$-complete graph) interaction rule; when $N_1\neq N_2$, no single value of $\alpha$ can eliminate size-induced asymmetries in the frequencies of within- and between-clique interactions, so the standard model cannot be recovered in this sense.
Conditional on selecting a node in clique~1, the probability that its interaction partner lies in clique~1 versus clique~2 is proportional to $\alpha(N_1-1)$ and $(1-\alpha)N_2$, respectively.
In Figure \ref{fig:networkexample}, intra-clique interactions occur more frequently than inter-clique interactions, corresponding to a large value of $\alpha$.
Our model is equivalent to that studied in \cite{gastner2019voter}, which is parameterized by the number of inter-clique connections $X$ and the fraction $r$ of nodes in the first clique. In our notation, $X=(1-\alpha)N_1N_2$ and $r=N_1/N$. While the interaction rules coincide, our analysis proceeds along different lines.

\subsection{Stochastic Approximation}
To analyze the discrete dynamics, we formulate a diffusion (stochastic differential equation, SDE) approximation for $x_1$ and $x_2$ by computing first and second moments of their increments over a single update.
Specifically, we compute the joint transition probabilities $p_{i,j}$ for the changes in $(x_1,x_2)$ over a time step $\Delta t = 2/N$, where $i,j \in \{-1,0,1\}$. Here, $p_{i,j}$ denotes the probability that $x_1$ and $x_2$ change by $i/N_1$ and $j/N_2$, respectively, during one update:
\begin{equation*}
     p_{i,j} := \mathbb{P}\left(x_1 \mapsto x_1+\frac{i}{N_1}, \; x_2 \mapsto x_2+\frac{j}{N_2}\right).
\end{equation*}
Each transition probability is obtained by conditioning on the type of interacting pair selected at the update step. We first illustrate this procedure with representative update events, characterized by changes $(i,j)$ in the number of $-1$ opinions in each clique.

\begin{figure}[t!]
    \centering
\includegraphics[width=0.7\linewidth]{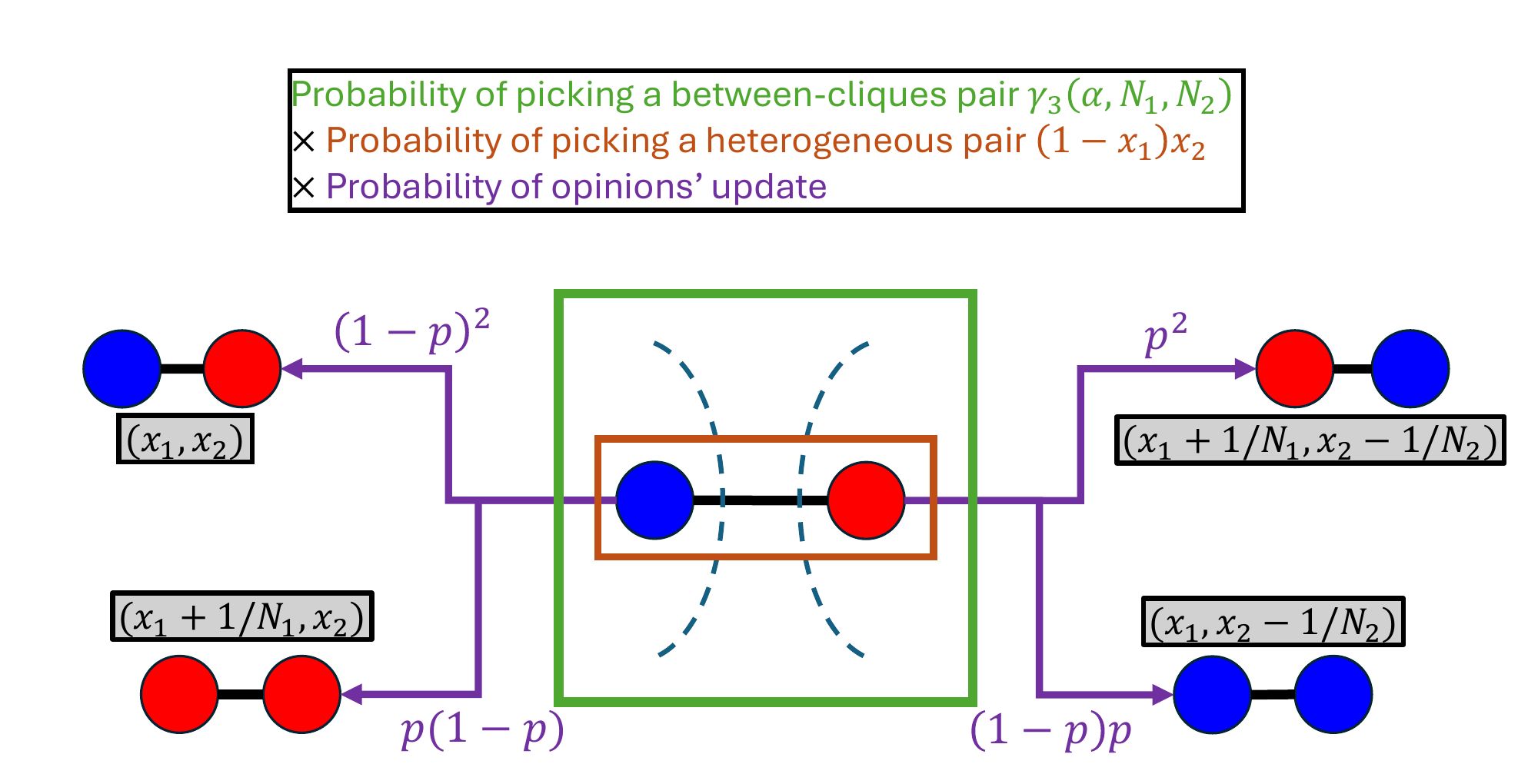}
    \caption{{\bf Example of an update event.} We illustrate the update of a heterogeneous $(+1,-1)$ pair (shown as blue and red nodes in cliques 1 and 2, respectively), starting from state $(x_1,x_2)$. An edge is first selected (green text and box) with dashed arcs indicating a between-clique interaction. A specific pair type is then chosen (brown text and box) with edges represented by black lines between nodes. Finally, opinions are updated probabilistically (indicated by purple text and an arrow), yielding the new state, which is highlighted in gray.}
    \label{fig:updateexample}
\end{figure}

Since both cliques cannot simultaneously lose or gain a $-1$ opinion, we have  $p_{\pm 1, \pm 1}=0$. The transition $(i,j) = (-1,0)$ occurs when a heterogeneous edge of type $(-1,+1)$ is selected within clique~1 or between cliques and only the $-1$ node flips and the $+1$ opinion remains unchanged, so that
\begin{equation*}
    p_{-1,0} = 2 \gamma_1 p(1-p) x_1 (1-x_1) + \gamma_3 p(1-p) x_1 (1-x_2).
\end{equation*}
The transition $(i,j) = (-1,+1)$ occurs when a heterogeneous inter-clique edge of type $(-1,+1)$ is selected and both opinions flip, yielding
\begin{equation*}
    p_{-1,+1} = \gamma_3 x_1 (1-x_2) p^2.
\end{equation*}
The remaining transition probabilities follow from analogous considerations; all nine are collected in Table~\ref{tab:prob}.
Figure \ref{fig:updateexample} provides a representative visual illustration of how specific entries arise.

\begin{table}[t!]
\centering
\begin{tabular}{|C{2.5cm}|C{4cm}|C{4cm}|C{4cm}|}
    \hline
     \cellcolor{gray!50} Update & \cellcolor{gray!25}  $x_2\to x_2-1/N_2$ &\cellcolor{gray!25} $x_2\to x_2$&\cellcolor{gray!25} $x_2\to x_2+1/N_2$\\
    \hline\cellcolor{gray!25} 
$x_1\to x_1-1/N_1$
    & $0$&$\gamma_1  2 x_1(1-x_1) p(1-p) + \gamma_3  x_1(1-x_2)p(1-p)$&$\gamma_3  x_1(1-x_2)p^2$ \\
    \hline\cellcolor{gray!25} 
    $x_1\to x_1$ & $\gamma_2  2 x_2(1-x_2) p(1-p) + \gamma_3  (1-x_1) x_2p(1-p) $&Complement &$\gamma_2  2 x_2(1-x_2) p(1-p) + \gamma_3  x_1(1-x_2)p(1-p)$ \\
    \hline\cellcolor{gray!25} 
    $x_1\to x_1+1/N_1$ & $\gamma_3  (1-x_1) x_2p^2$ & $ \gamma_1  2 x_1(1-x_1) p(1-p) + \gamma_3  (1-x_1) x_2p(1-p)$ & $0$ \\
    \hline
\end{tabular}
    \caption{\textbf{Joint transition probabilities for a single update.} Rows: change in $x_1$; columns: change in $x_2$. ``Complement'' denotes $p_{0,0} = 1 - \sum_{(i,j)\neq(0,0)} p_{i,j}$. The time step is $\Delta t=2/N$.}
    \label{tab:prob}
\end{table}

Using the joint transition probabilities $p_{i,j}$, we compute the moments of the increments $\Delta x_i:=x_i(t+\Delta t)-x_i(t)$. The expected change in $x_1$ over one update is
\begin{equation*}
\begin{split}
        \Exp(\Delta x_1) & = \frac{1}{N_1} \sum_{i,j} i p_{i,j}=\frac{\gamma_3p}{N_1}\Big(x_2-x_1\Big),
\end{split}
\end{equation*}
and similarly,
\begin{equation*}
    \begin{split}
        \Exp(\Delta x_2) & = \frac{1}{N_2} \sum_{i,j} jp_{i,j}=\frac{\gamma_3p}{N_2}\Big(x_1-x_2\Big).
    \end{split}
\end{equation*}
Moreover, the second moments satisfy
\begin{equation*}
\begin{split}
        \Exp(\Delta x_1^2) = \frac{1}{N_1^2}\sum_{i,j} i^2\,p_{i,j}
        =\frac{1}{N_1^2}\Big((\gamma_1  4 x_1(1-x_1) p(1-p)+\gamma_3p(x_1+x_2-2x_1x_2)\Big),
\end{split}
\end{equation*}
and
\begin{equation*}
\begin{split}
        \Exp(\Delta x_2^2)
        = \frac{1}{N_2^2}\sum_{i,j} j^2\,p_{i,j}
        =\frac{1}{N_2^2}\Big((\gamma_2  4 x_2(1-x_2) p(1-p)+\gamma_3p(x_1+x_2-2x_1x_2)\big),
\end{split}
\end{equation*}
and the mixed second moment
\begin{equation*}
\begin{split}
        \Exp(\Delta x_1\Delta x_2)
        = \frac{1}{N_1N_2}\sum_{i,j} ij\,p_{i,j}
        =-\frac{1}{N_1N_2}\Big(\gamma_3p^2(x_1+x_2-2x_1x_2)\Big).
\end{split}
\end{equation*}
Finally, we define the variances and covariance of the increments via
\begin{equation*}
\Var(\Delta x_i) := \Exp(\Delta x_i^2) - \Exp(\Delta x_i)^2,
\qquad
\Cov(\Delta x_1,\Delta x_2) := \Exp(\Delta x_1\Delta x_2)
- \Exp(\Delta x_1)\Exp(\Delta x_2).
\end{equation*}
Using the expressions derived above, this yields
\begin{equation*}
\begin{split}
        \Var(\Delta x_1)
        &=\frac{1}{N_1^2}\Big((\gamma_1  4 x_1(1-x_1) p(1-p)+\gamma_3p(x_1+x_2-2x_1x_2)-\gamma_3^2p^2(x_2-x_1)^2\Big),
\end{split}
\end{equation*}
\begin{equation*}
\begin{split}
       \Var(\Delta x_2)
        &=\frac{1}{N_2^2}\Big((\gamma_2  4 x_2(1-x_2) p(1-p)+\gamma_3p(x_1+x_2-2x_1x_2)-\gamma_3^2p^2(x_1-x_2)^2\Big),
\end{split}
\end{equation*}
and
\begin{equation*}
\begin{split}
        \Cov(\Delta x_1,\Delta x_2)&=\frac{1}{N_1N_2}\Big(-\gamma_3p^2(x_1+x_2-2x_1x_2)+\gamma_3^2p^2(x_2-x_1)^2\Big).
\end{split}
\end{equation*}

Let $\Sigma(\mathbf{x})$ denote the covariance matrix of the increment
$\Delta\mathbf{x} := (\Delta x_1,\Delta x_2)^\top$, with entries defined above. We introduce a diffusion matrix $\sigma (\mathbf{x})$ satisfying
$\Sigma = \sigma\sigma^\top$, noting that $\sigma (\mathbf{x})$ is not unique.

Over a macroscopic time interval $\Delta t=1$, the dynamics consist of $N/2$ update events, so that on average each agent participates in one interaction.
Rescaling time accordingly and retaining the leading-order drift and diffusion terms yields an SDE approximation of the discrete voter dynamics according to a two-dimensional Ornstein-Uhlenbeck process~\cite{gardiner1985,ethier1986markov,vankampen2007stochastic}:
\begin{equation}\label{eq:SDEs}
    \begin{cases}
 dx_1= \displaystyle \frac{N}{2} \, \Exp(\Delta x_1) \,dt+ \sqrt{\frac{N}{2}}\,\bigl(\sigma_{11}\,dW_1 + \sigma_{12}\,dW_2\bigr),\\
    dx_2=\displaystyle \frac{N}{2}\,\Exp(\Delta x_2)\,dt
      + \sqrt{\frac{N}{2}}\,\bigl(\sigma_{21}\,dW_1 + \sigma_{22}\,dW_2\bigr),
    \end{cases}
\end{equation}
where $(W_1,W_2)$ is a two-dimensional Wiener process.
The state variables are constrained to the unit interval, $0\le x_i\le 1$, for $i=1,2$.

A major consequence of the diffusion approximation is the need to redefine consensus. For the associated stochastic process to exit the domain with positive probability, the absorbing portion of the boundary must have positive measure. Consequently, consensus cannot be represented solely by the two corner points corresponding to complete agreement.
We therefore enlarge the absorbing boundary by introducing a tolerance parameter $\theta> 0$. We say that a {\it consensus} is reached if either
\begin{equation*}
    \norm{(N_1x_1,N_2x_2)}\leq \theta N \quad\text{OR}\quad\norm{(N_1(1-x_1),N_2(1-x_2))}\leq \theta N,
\end{equation*}
where $\norm{\cdot}$ denotes a chosen norm. Throughout this paper, we take $\norm{\cdot} = \norm{\cdot}_1$. This definition corresponds to a state in which at most a fraction $\theta$ of dissenters disagree with the majority.

For numerical simulations, we use $0.01\leq \theta\leq 0.025$. In real voting settings, social rules requiring more than half of participants to agree are referred to as \textit{majority rule}, while larger thresholds (\textit{e.g.} three-quarters) are termed \textit{supermajority} rules~\cite{mcgann2004tyranny}. The values of $\theta$ considered here correspond to thresholds much closer to unanimity; informally, they could be described as an \textit{uber-/ultra-majority}. For consistency, we refer to this condition as consensus throughout.

\subsection{The Continuum Approximation}
\label{sec:cont}
Applying the functional central limit theorem yields the PDE formulation. Let $P(\xx,t)$ represent the probability of finding the process at a given state $\xx = (x_1, x_2)$ at time $t$. More precisely, $P$ is a probability density function of an associated random variable $(\xx,t)$. The system \eqref{eq:SDEs} thus has a Fokker-Planck representation \cite{risken1989fokker}:
\begin{equation}\label{eq:fwdFP}
\frac{\partial P}{\partial t} (\xx,t)
= -\nabla \cdot \big(\boldsymbol{\mu}(\mathbf{x})\,P (\xx,t)\big)
+ \nabla \cdot \nabla \cdot \big(D(\mathbf{x})\,P (\xx,t) \big),
\end{equation}
where $\boldsymbol{\mu}(\mathbf{x})=(\mu_1,\mu_2)$
is the drift vector with
\begin{align*}
\mu_i(\mathbf{x}) := \Exp(\Delta x_i)\,\frac{N}{2},
\end{align*}
and $D(\mathbf{x})$ is the diffusion matrix defined by
\begin{align*}
D(\mathbf{x}) := \sigma(\mathbf{x})\sigma(\mathbf{x})^\top\,\frac{N}{4}
= \mathbf{Cov}(\Delta x_1,\Delta x_2)\,\frac{N}{4}.
\end{align*}
The spatial domain is $[0,1]^2$ with the consensus corners removed, and we denote it by $\Omega^\theta$.
By the Cauchy-Schwarz inequality applied to the variance--covariance relation,
the operator in \eqref{eq:fwdFP} is parabolic in the interior of $\Omega^\theta$.

We assume a point initial condition for the SDE $(x_1(0), x_2(0)) = (y_1,y_2) \equiv \yy$, as described by
\begin{align*}
P_{\text{in}}(\xx):=P(\xx,0) &=\delta(\xx - \yy),
\end{align*}
which will allow us to specify the associated backward equation for the first-passage time statistics. It is convenient to write \eqref{eq:fwdFP} in conservation (divergence) form,
\begin{align}\label{eq:fwdFPdiv}
\frac{\partial P}{\partial t} +\nabla\cdot \mathbf{J} &=0, \ \ \ \mathbf{J}(\mathbf{x},t) :=\boldsymbol{\mu}(\mathbf{x})\,P(\mathbf{x},t) -\nabla\cdot\!\big(D(\mathbf{x})\,P(\mathbf{x},t)\big),
\end{align}
where
\begin{align*}
\big(\nabla\cdot(DP)\big)_i &:=\sum_{j=1}^2 \frac{\partial}{\partial x_j} \big(D_{ij}(\mathbf{x})\,P(\mathbf{x},t)\big), \qquad i=1,2.
\end{align*}
We decompose the boundary of $\Omega^\theta$ as
\begin{align*}
\partial\Omega^\theta &=\partial\Omega^\theta_{\mathrm{ref}} \cup \partial\Omega^\theta_{\mathrm{abs}},\\
\partial\Omega^\theta_{\mathrm{abs}} &:=\Gamma_0^\theta\cup\Gamma_1^\theta,\\
\partial\Omega^\theta_{\mathrm{ref}} &:=\partial\Omega^\theta\setminus \partial\Omega^\theta_{\mathrm{abs}},
\end{align*}
with absorbing (homogeneous Dirichlet) boundary segments
\begin{align*}
\Gamma_0^\theta &:=\Big\{(x_1,x_2)\in[0,1]^2:\; N_1x_1+N_2x_2=\theta N\Big\},\\
\Gamma_1^\theta &:=\Big\{(x_1,x_2)\in[0,1]^2:\; N_1(1-x_1)+N_2(1-x_2)=\theta N\Big\}.
\end{align*}
We impose reflecting (no-flux) boundary conditions on $\partial\Omega^\theta_{\mathrm{ref}}$,
\begin{align*}
\mathbf{J} (\xx,t) \cdot\mathbf{n} &=0 \qquad \text{for} \ \xx \in \partial\Omega^\theta_{\mathrm{ref}},
\end{align*}
and absorbing boundary conditions on $\partial\Omega^\theta_{\mathrm{abs}}$,
\begin{align*}
P (\xx,t) &=0 \qquad \text{for} \ \xx \in  \partial\Omega^\theta_{\mathrm{abs}},
\end{align*}
where $\mathbf{n}$ denotes the outward unit normal on $\partial\Omega^\theta$.

Given the PDE formulation \eqref{eq:fwdFP}, we now derive the mean consensus time.
For this purpose, we regard the solution $P(\xx,t)$ as a function of the
initial state $\mathbf{y}=(y_1,y_2)$ through the point initial condition
$P(\xx,0)=\delta(\xx - \yy)$, and write
\begin{align*}
P(\xx,t) = P(\xx,t \mid \mathbf{y},0).
\end{align*}
In this formulation, $\mathbf{y}$ serves as the state variable in the associated
backward problem for first-passage time statistics.
We define a corresponding survival probability
\begin{align*}
G(\yy,t) &:= \int_{\Omega^\theta} P(\xx,t \mid \yy,0)\,d \xx , 
\end{align*}
which represents the probability that the process has not yet reached the
absorbing boundary by time $t$, given the initial state $\yy$.

Let $T(\yy)$ denote the first exit time of the process from $\Omega^{\theta}$. Then,
\begin{equation*}
    \Prob(T(\yy)\geq t)=G(\yy,t).
\end{equation*}

The adjoint of the operator associated with \eqref{eq:fwdFP} defines the backward Kolmogorov equation for functions of the initial state $\mathbf{y} = (y_1, y_2)$,
\begin{equation}\label{eq:bwdFP}
\frac{\partial P}{\partial t} (\yy,t) = \boldsymbol{\mu}(\mathbf{y})\cdot\nabla_{\mathbf{y}} P (\yy,t) + \nabla_{\mathbf{y}}\cdot\nabla_{\mathbf{y}}\cdot \big(D(\mathbf{y})\,P (\yy,t) \big),
\end{equation}
where the drift and diffusion coefficients are defined as in \eqref{eq:fwdFP}, but evaluated at $\yy$ instead of $\mathbf{x}$. Integrating \eqref{eq:bwdFP} over the domain $\Omega^\theta$ yields the corresponding backward equation for the survival probability $G$.
\begin{equation}\label{eq:bwdFPforG}
\frac{\partial G}{\partial t} (\yy,t) = \boldsymbol{\mu}(\mathbf{y})\cdot\nabla_{\mathbf{y}} G(\yy,t) + \nabla_{\mathbf{y}}\cdot\nabla_{\mathbf{y}}\cdot \big(D(\mathbf{y})\,G (\yy,t) \big).
\end{equation}
The function $G(\yy,t)$ is the survival probability associated with the
first exit (consensus) time. Its negative derivative gives the probability density function of the first exit time. Using standard first-passage arguments~\cite{gardiner1985} together with the no-flux boundary conditions \eqref{eq:fwdFPdiv}, the mean consensus time satisfies
\begin{equation*}
    T(\yy)=\int_0^\infty G(\yy,t) dt.
\end{equation*}
Since consensus is always reached, $G(\yy,t)\to 0$ as $t\to\infty$ and $G(\yy,0)=1$, hence
$\int_0^\infty \partial_t G(\yy,t)\,dt = G(\yy,\infty)-G(\yy,0) = -1$.
Integrating \eqref{eq:bwdFPforG} with respect to time then yields
\begin{equation}\label{eq:PDEforT}
\begin{split}
        -1=&\mu_1(\yy)\frac{\partial T}{\partial y_1}+\mu_2(\yy)\frac{\partial T}{\partial y_2}\\
        &+D_{11}(\yy)\frac{\partial^2 T}{\partial y^2_1}
        +2D_{12}(\yy)\frac{\partial^2 T}{\partial y_1\partial y_2}+D_{22}(\yy)\frac{\partial^2 T}{\partial y^2_2}.
\end{split}
\end{equation}
Now we comment on the boundary conditions. We still have no-flux boundary conditions dictated by the derivation of \eqref{eq:bwdFP}. However, there is a caveat regarding the form of the flux, as seen in the following divergence form of \eqref{eq:PDEforT}
\begin{equation}\label{eq:PDEforTdiv}
    \begin{split}
        &1+\nabla\cdot \big[D (\mathbf{y}) \nabla T \big]+\begin{bmatrix}
            -\frac{\partial D_{11}}{\partial y_1}-\frac{\partial D_{12}}{\partial y_2}+\mu_1 (\mathbf{y})\\
            -\frac{\partial D_{12}}{\partial y_1}-\frac{\partial D_{22}}{\partial y_2}+\mu_2 (\mathbf{y})
        \end{bmatrix}\cdot\nabla T=0.
    \end{split}
\end{equation}
The term after $\nabla\cdot$ in \eqref{eq:PDEforTdiv} is the ``backward" flux for \eqref{eq:PDEforT}. When comparing fluxes in \eqref{eq:fwdFPdiv} and \eqref{eq:PDEforTdiv}, in addition to sign changes, the drift terms (those that include $\mu_i$) are absorbed into the first-order (reaction) term. Instead, the drift terms and higher derivatives of diffusion rates are contained in the non-divergence contribution to the operator.

Having established the backward problem for the mean consensus time and clarified the associated boundary conditions, we now turn to the forward equation \eqref{eq:fwdFP} to understand how probability mass is transported through the state space. In particular, we exploit the separation between $\mathcal{O}(1)$ drift terms and $\mathcal{O}(1/N)$ diffusive terms to identify the dominant dynamics and the relevant boundary-layer scalings in the large-population limit.

\section{Multiscale Asymptotics of Consensus Formation}
In this section, we analyze \eqref{eq:fwdFP} using asymptotic methods in the limit of large populations $N\to\infty$, to identify the mechanisms governing consensus formation across distinct time scales. Extreme values of other parameters (\textit{e.g.}, $p=0$ or $p=1$) lead to degenerate dynamics and singular behavior; these cases are examined numerically in the next section, while their analytical treatment is left for future work.  Throughout this section, we assume equal population sizes, $N_1=N_2=N/2$. This symmetry simplifies both the equations and the domain, making the analysis more tractable. Nevertheless, the same approach can be extended in the case that $N_1\not=N_2$, provided both $N_i\to\infty$.  

First, we introduce the change of variables
\begin{equation*}
    w=\frac{x_1+x_2-1}{\sqrt{2}}\;\text{and}\; z=\frac{x_2-x_1}{\sqrt{2}}.
\end{equation*}
This transformation centers the domain at $(0,0)$ and aligns the coordinate axes with the diagonal and anti-diagonal directions, mapping the original diagonal $x_1=x_2$, corresponding to alignment of the belief fractions in each module, onto the horizontal axis. The variable $w$ runs from $-(1-2\theta)/\sqrt{2}$ to $(1-2\theta)/\sqrt{2}$, while $z$ runs from $-1/\sqrt{2}$ to $1/\sqrt{2}$. Recall that the domain is a truncated diamond whose left- and right-hand corners have been removed. In the new variables, the drift and diffusion terms naturally separate along the diagonal ($w$) and transverse ($z$) directions, so that equation \eqref{eq:fwdFP} becomes 
\begin{equation}\label{eq:fwdFPinzw}
    \begin{split}
      \frac{\partial P}{\partial t}=&2p(1-\alpha) \frac{\partial}{\partial z}[zP]+\frac{p(1-p)}{2N} \left[ \frac{\partial^2}{\partial w^2} + \frac{\partial^2}{\partial z^2} \right] \Big[(1-2w^2+2(1-2\alpha)z^2)P\Big]\\
      & \qquad +\frac{p^2(1-\alpha)}{N} \frac{\partial^2}{\partial z^2} \Big[(1-2w^2-2(1-2\alpha)z^2)P\Big]-\frac{4\alpha p(1-p)}{N} \frac{\partial^2}{\partial w \partial z} \Big[wzP\Big],
    \end{split}
\end{equation}
where we have expanded the coefficients to make their dependence on $w$ and $z$ explicit.

In the singular limit $N\to\infty$, all second-derivative terms vanish and
\eqref{eq:fwdFPinzw} reduces to the first-order transport equation
\begin{align} \label{eq:zeroth1}
\frac{\partial P^0}{\partial t} (w,z,t) =2p(1-\alpha) \frac{\partial}{\partial z} [zP^0(w,z,t)].
\end{align}
Along characteristics, $w$ is constant and $z(t)=z_0e^{-2p(1-\alpha)t}$. Hence, if
$P^0(w,z,0)=P_{\rm in}(w,z)$, then
\begin{equation}\label{eq:zerothsol1}
P^0(w,z,t)=e^{2p(1-\alpha)t}\,P_{\rm in}\!\left(w,ze^{2p(1-\alpha)t}\right)=\delta(w-w_0)\delta(z-z_0e^{-2p(1-\alpha)t}).
\end{equation}
This implies exponential contraction toward the diagonal $z=0$, with deviations transverse to alignment decaying at rate $2p(1-\alpha)$ (equivalently, on the characteristic timescale $[2p(1-\alpha)]^{-1}$).

As a result, probability mass rapidly concentrates near $z=0$ on $\mathcal{O}(1)$ timescales. Diffusive effects, which are suppressed by a factor of $1/N$ relative to the drift, become relevant only within a narrow boundary layer around $z=0$.
This separation of scales motivates a multiscale expansion in $\epsilon=1/N$,
\begin{equation*}
P(w,z,t) = P^0(w,z,t) + \epsilon P^1(w,z,t) + \epsilon^2 P^2(w,z,t) + \cdots,
\end{equation*}
where $P^0$ captures the rapid drift toward alignment and higher-order terms encode finite-size diffusive corrections.

To describe the subsequent evolution once the distribution has concentrated near the alignment manifold $z=0$, we examine an inner region where diffusive effects compete with the vanishing drift. 
This motivates the introduction of rescaled spatial and temporal variables that capture the dynamics within a boundary layer around $z=0$.

\subsection{Boundary-Layer Dynamics Near the Alignment Manifold}

When $z$ becomes small ($z = \mathcal{O}(\sqrt{z}$), we must resolve dynamics by introducing a stretched variable  $s=z/\sqrt{\epsilon}$.
In the inner region, the probability density therefore depends on $(w,s,t,T)$.
At leading order, the dynamics are confined to the $z$-direction, with no evolution in $w$; evolution in $w$ occurs on a slower timescale.

\begin{figure}[t!]
    \begin{center} \includegraphics[width=16cm]{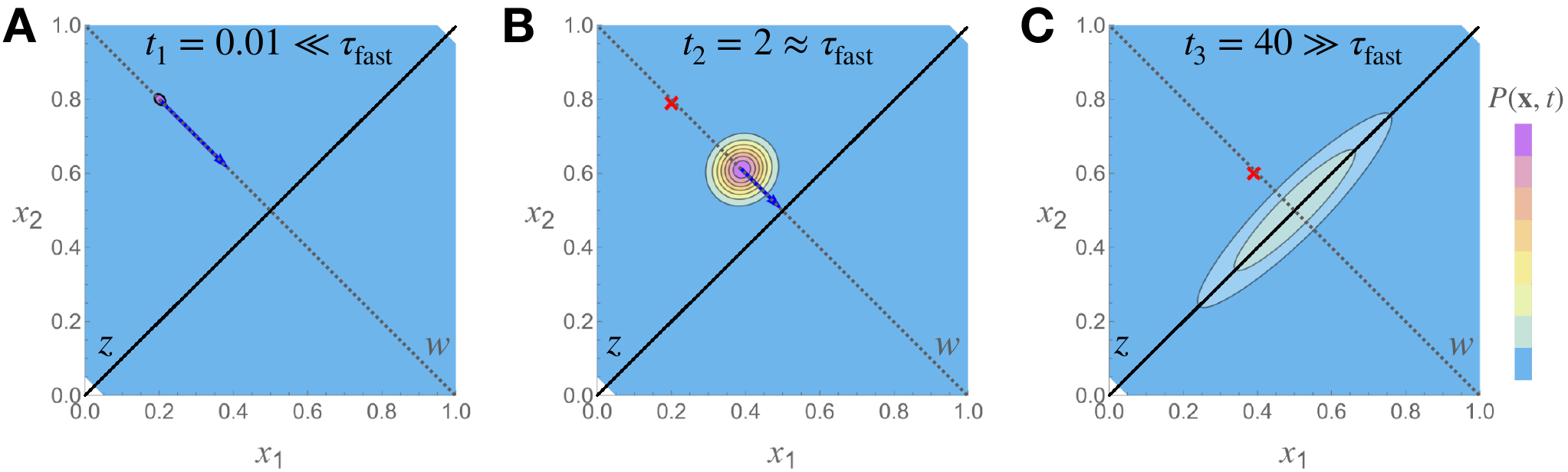} \end{center}
    \vspace{-6mm}
    \caption{\textbf{Separation of fast and slow dynamics in the voter model.} The solid line is the diagonal $z=0$; dashed lines show the rotated coordinates $w$, $z$. Red crosses and blue arrows track successive centres of mass. \textbf{(A)}~For $t \ll \tau_{\mathrm{fast}} = [2p(1-\alpha)]^{-1}$, the density drifts toward the diagonal as predicted by~\eqref{eq:zerothsol1}. \textbf{(B)}~At $t \approx \tau_{\mathrm{fast}}$, diffusion competes with the weakening drift, producing the Gaussian profile~\eqref{eq:zerothsol2}. \textbf{(C)}~For $t \gg \tau_{\mathrm{fast}}$, the density is narrow in $z$ (width $\propto \sqrt{\epsilon(1-2w^2)}$) and elongated along the diagonal, reflecting slow $w$-dynamics on the $T=\epsilon\,t$ timescale. Peak reduction from \textbf{(B)} to \textbf{(C)} reflects absorbing-boundary mass loss. Parameters: $N_1 = N_2 = 100$, $\alpha = 0.5$, $p = 0.5$, $\theta = 0.025$; initial condition $\mathbf{x}_0 = (1/5,\, 4/5)$.}
    \label{fig:bellevol}
\end{figure}

Let $s_0=\mathcal{O}(1)$. We first shift time by
$$t_{in}:=\frac{1}{2p(1-\alpha)}\log \frac{z_0}{s_0\sqrt{\epsilon}},$$
which corresponds to the time at which $z=s_0\sqrt{\epsilon}$.
We then introduce the slow time scale $T=\epsilon t$, treated as an independent 
variable in the inner region.
Accordingly, in the inner region, the density is written as $P(w,s,t,T)$.
With these rescalings, the leading-order term satisfies
\begin{equation}\label{eq:zeroth2}
    \frac{\partial P^0}{\partial t}(w,s,t,T)=2p(1-\alpha) \frac{\partial}{\partial s}\left[ s P^0 \right]  +\frac{p+p^2-2\alpha p^2}{2}(1-2w^2)\frac{\partial^2 P^0}{\partial s^2}.
\end{equation}
Equation \eqref{eq:zeroth2} differs from the outer transport equation \eqref{eq:zeroth1} by the presence of an additional $w$-dependent diffusion term.
The initial condition is
\begin{align*}
    P(w,s,0,0) = \delta(w-w_0)\,\delta(\sqrt{\epsilon}(s-s_0)).
\end{align*}
Here ``initial" refers to the shifted variables $t'=t-t_{in}$ and $T'=\epsilon(t-t_{in})$, so that $t'=0$ describes the moment the solution enters the boundary layer; the initial condition is imposed at this moment.
Solving \eqref{eq:zeroth2} with this initial condition yields
\begin{equation} \label{eq:zerothsol2}
    \begin{split}
        P^0&=\frac{A(w,T')}{\sqrt{4\pi\epsilon\frac{p+p^2-2\alpha p^2}{8p(1-\alpha)}(1-e^{-4p(1-\alpha)t'})(1-2w^2)}}\exp\left[-\frac{(z-s_0\sqrt{\epsilon}e^{-2p(1-\alpha)t'})^2}{4\epsilon\frac{p+p^2-2\alpha p^2}{8p(1-\alpha)}(1-e^{-4p(1-\alpha)t'})(1-2w^2)}\right]\\
        &\equiv A(w,T')B(s,t'),
    \end{split}
\end{equation}
where $B(z,t')$ denotes the Gaussian factor, with mean $s_0\sqrt{\epsilon}\,e^{-2p(1-\alpha)t'}$ and variance $2\epsilon\,\frac{p+p^2-2\alpha p^2}{8p(1-\alpha)}(1-e^{-4p(1-\alpha)t'})(1-2w^2)$.
Equation \eqref{eq:zerothsol2} corresponds to a modified Ornstein--Uhlenbeck--type 
solution~\cite{pavliotis2014stochastic}: compared with \eqref{eq:zerothsol1}, it introduces diffusive smoothing, though the 
overall dynamics are unchanged. Probability mass approaches the alignment manifold $z=0$, 
advection diminishes, and diffusion takes over.
$P^0$ is the product of $A(w,T')$ and the following Gaussian density in $z$
\begin{equation}\label{eq:NormalDis}
    \mathcal{N}\left(s_0\sqrt{\epsilon}e^{-2p(1-\alpha)t'},\epsilon\frac{p+p^2-2\alpha p^2}{4p(1-\alpha)}(1-e^{-4p(1-\alpha)t'})(1-2w^2)\right).
\end{equation}
This shows that the dynamics in the $z$-direction follow a Gaussian density centered at $s_0\sqrt{\epsilon}\,e^{-2p(1-\alpha)t'}$, which decays to zero as $t'\to\infty$. Its width grows from zero and saturates at a level proportional to $\sqrt{1-2w^2}$, vanishing as $w\to\pm 1/\sqrt{2}$.
 Note that since we ``cut" the corners, we do not run into singularities. The speed at which mass concentrates at $z=0$ is still dependent on $p(1-\alpha)$, similar to \eqref{eq:zerothsol1}. The presence of $\epsilon$ is an artifact of the initial condition that changed when $z\to s$. As $\epsilon$ appears in both the mean and variance, this signifies how population size affects the spread: the larger the population, the closer to zero and the thinner the Gaussian ``bell" in $z$. The simulation of the original equation \eqref{eq:fwdFP} is shown in Figure \ref{fig:bellevol}. The numerical results, even for relatively small populations, confirm our preliminary findings on the dynamics. In Figure \ref{fig:bellevol}{\bf A}, after a relatively small time, we notice that the initial mass spreads as it approaches the main diagonal along the anti-diagonal. In Figure \ref{fig:bellevol}{\bf B}, after 2 time units, the movement towards the diagonal continues. The spread is more apparent and appears to be equal in both directions. Finally, in Figure \ref{fig:bellevol}{\bf C}, the spread primarily happens along the main diagonal, while it remains thin along the anti-diagonal. The slow dynamics take effect. The solution ``leaks" through the absorbing boundaries, hence the depletion of the maximum values and overall mass. 

\begin{figure}[t!]
\begin{center} \includegraphics[width=15cm]{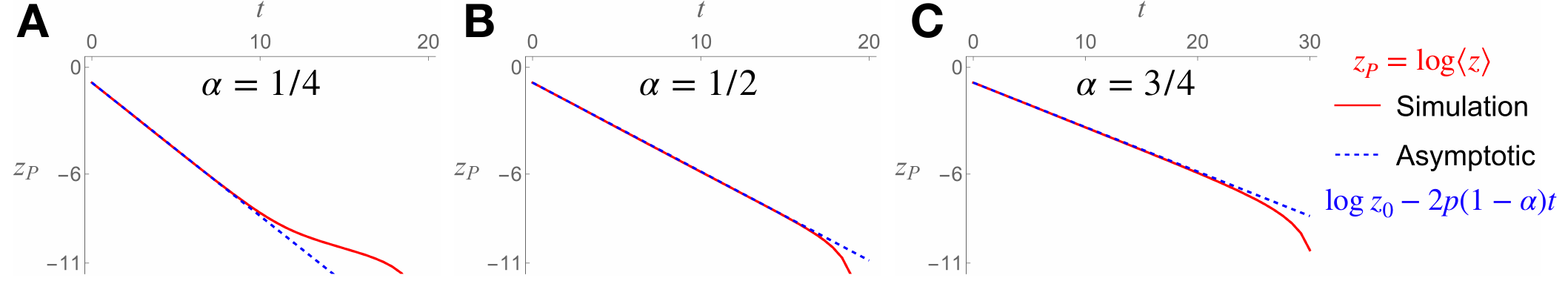} \end{center}
    \caption{\textbf{Accuracy of asymptotic approximation for the fast dynamics.} The red curve shows $z_P = \log\langle z \rangle$, the natural log of the mean $z$-coordinate of $P$, computed from numerical simulation of~\eqref{eq:fwdFP} using a center-of-mass formula based on $10^5$ sample points. The blue dashed curve is the asymptotic prediction $\log z_0 - 2p(1-\alpha)t$ from~\eqref{eq:zerothsol1}. The two agree well during the fast-drift regime and diverge once boundary-layer diffusion and absorbing-boundary effects become significant. Panels differ in $\alpha$: \textbf{(A)}~$\alpha=1/4$, \textbf{(B)}~$\alpha=1/2$, \textbf{(C)}~$\alpha=3/4$. Remaining parameters: $y_1=1/5$, $y_2=4/5$, $N_1=N_2=100$, $p=1/2$, $\theta=0.025$.}
    \label{fig:asymp}
\end{figure}

Figure \ref{fig:asymp} showcases how the asymptotic result agrees with numerical simulations of the actual PDE \eqref{eq:fwdFP} across various parameter combinations. We start with a point far from the diagonal and track how long it takes for the point to reach the diagonal. We observe that the agreement holds until $z$ gets considerably small, with the threshold varying with $\alpha$. Hence, increasing $\alpha$ extends the window of agreement. This can be explained by the appearance of $\alpha$ in the variance of \eqref{eq:NormalDis}: the larger $\alpha$ is, the narrower the Gaussian. In other words, the solution preserves its shape, dictated by the initial conditions, longer with larger $\alpha$. Beyond the threshold, the two functions diverge: the numerical $z$-coordinate still approaches zero, but at a sub-exponential rate, visible as the red 
curve bending above the asymptotic straight line in the $\log$ plot. The exact divergence threshold is not known analytically and likely depends on the interplay between boundary-layer diffusion and the absorbing boundaries. Nevertheless, Figure \ref{fig:asymp} confirms that the leading-order asymptotic expansion accurately captures the fast dynamics across a range of $\alpha$ values.

We now proceed to the next order in the expansion. Dropping primes from the shifted temporal variables gives
\begin{equation}\label{eq:firstordcond}
    \begin{split}
            &\frac{\partial P^1}{\partial t}-2p(1-\alpha) \frac{\partial}{\partial s}\left[s P^1\right]-\frac{p+p^2-2\alpha p^2}{2}(1-2w^2)\frac{\partial^2 P^1}{\partial s^2}\\
            &\quad=\big[C_4+4C_2(L-1)-4C_3L\big] P^0
            +\big[-8C_2+C_4\big]\frac{\partial P^0}{\partial w}+\big[s(C_4+8L(C_2-C_3))\big] \frac{\partial P^0}{\partial s}\\
            &\quad+C_2(1-2w^2)\frac{\partial ^2 P^0}{\partial w^2}+2 Ls^2 (C_2-C_3)\frac{\partial^2 P^0}{\partial s^2}+C_4ws\frac{\partial^2 P^0}{\partial w \partial s}-\frac{\partial P^0}{\partial T},
    \end{split}
\end{equation}
where
\begin{equation}\label{eq:coeffP1}
    \begin{split}
        C_2=p(1-p)/2,\quad C_3=p^2(1-\alpha),\quad C_4=-4\alpha p(1-p),\quad L=2(1-2\alpha).
    \end{split}
\end{equation}
Note, the coefficients in \eqref{eq:coeffP1} coincide with those of the diffusion operator in \eqref{eq:fwdFP}; we retain them in this form since simplifications will follow. Equation \eqref{eq:firstordcond} has a solution in $P^1$ only if the right-hand side satisfies a Fredholm solvability condition: it must be orthogonal to the null space of the adjoint of the $s$-operator on the left-hand side, which is spanned by constants~\cite{pavliotis2014stochastic}. Hence, we integrate the right-hand side over the boundary-layer domain $\Omega_s^{\theta}$ in the stretched variable $s$ (equivalently $\Omega^\theta_z$ for $z$) and set the result to zero
\begin{equation*}
    \begin{split}
           \int_{\Omega^\theta_s} \frac{\partial P^0}{\partial T} ds&=\int_{\Omega^\theta_s} \Bigg\{ \big[C_4+4C_2(L-1)-4C_3L\big] P^0
            +\big[-8C_2+C_4\big]\frac{\partial P^0}{\partial w}\\
            &\quad+\big[s(C_4+8L(C_2-C_3))\big] \frac{\partial P^0}{\partial s}
            +C_2(1-2w^2)\frac{\partial^2 P^0}{\partial w^2}\\
            &\quad+2 Ls^2 (C_2-C_3)\frac{\partial^2 P^0}{\partial s^2}+C_4ws\frac{\partial ^2P^0}{\partial w \partial s} \Bigg\}ds.
    \end{split}
\end{equation*}
Since $P^0 = A(w, T)\,B(s,t)$, integrating over $s$ eliminates the fast variable and yields an evolution equation for $A(w,T)$. To evaluate the resulting integrals, we impose
\begin{equation*}
    \int_{\Omega^\theta_s}B(s,t)=1\quad\text{and}\quad s^nB(s,t)\big|_{\partial \Omega^\theta_s}=0\quad\text{for }n=0,1.
\end{equation*}
The first condition normalizes $B$ as a density of $s$; the second ensures that boundary terms vanish after integration by parts. Both are justified by the fact that $B$ is a narrow Gaussian in $s$ for fixed $w$, as established in \eqref{eq:NormalDis}.
With these assumptions and after a few integrations by parts, we have
\begin{equation}\label{eq:wTeq}
    \frac{\partial A}{\partial T}=C_2\left[-4A-8w\frac{\partial A}{\partial w}+(1-2w^2)\frac{\partial^2 A}{\partial w^2}\right],
\end{equation}
where $C_2$ is defined in \eqref{eq:coeffP1}. Separating variables as $\tau(T)\mathcal{W}(w)$ shows the temporal part satisfies $\tau(T)=e^{\lambda C_2 T}$.
Since $C_2$ only depends on $p(1-p)$, the consensus time in this regime is minimized when $p=1/2$, which we verify numerically in the next section. Moreover, $\alpha$ is absent from \eqref{eq:wTeq}, so modularity influences the fast alignment stage but not the slow dynamics along the aligned manifold. Introducing $u = (1 + \sqrt{2}w)/2$, the spatial part reduces to
\begin{equation*}
    u(1-u)\frac{d^2\mathcal W}{du^2}
    +2(1-2u)\frac{d\mathcal W}{du}
    +\mu \mathcal W = 0 \qquad \text{with} \qquad \mu=-\frac{4+\lambda}{2}.
\end{equation*}
Equivalently, setting $v=\sqrt{2}w\in[-1,1]$ yields \begin{equation*}
(1-v^2)\frac{d^2\mathcal W}{dv^2} -4v\frac{d\mathcal W}{dv} +\mu \mathcal W =0. 
\end{equation*}
This is the Jacobi differential equation with parameters $(\alpha,\beta)=(1,1)$, whose eigenfunctions are the Jacobi polynomials $\mathcal P_n^{(1,1)}(v)$ and eigenvalues $\mu_n=n(n+3)$, $n\in\mathbb N\cup\{0\}$. Therefore, \[ \lambda_n=-2n(n+3)-4=-2n^2-6n-4. \] Since all eigenvalues are negative, every mode decays in time, confirming that consensus is reached with probability one. Thus, once the cliques have aligned, the slow dynamics are insensitive to modularity and reduce to the same one-dimensional neutral diffusion structure that underlies the well-mixed model.

To summarize: the multiscale analysis reveals two distinct stages of consensus formation. On $\mathcal{O}(1)$ timescales, inter-clique coupling drives exponential alignment of opinion fractions at rate $2p(1-\alpha)$. On the slow $T = t/N$ timescale, the system evolves along the alignment manifold via a diffusion whose rate depends on $p(1-p)$ but is independent of the modularity parameter $\alpha$. In the next section, we validate these predictions with numerical simulations.

\section{Consensus Times Across Parameter Regimes}
The asymptotic analysis of the previous section predicts that consensus evolves first by a fast alignment stage governed by $\alpha,$ followed by a slow diffusion stage governed by $p(1-p)$. However, it is not clear how these stages interact across parameter regimes, particularly when the symmetry assumptions underlying the analysis are relaxed.
In this section, we address this question numerically. We begin by verifying that the discrete, SDE, and PDE formulations agree across a range of parameters, then turn to the central question: is there an optimal level of inter-clique coupling that minimizes the time to reach consensus?
We will show that the answer depends critically on population imbalance and initial polarization, with nontrivial optima emerging precisely when the system is far from symmetric. The discrete and SDE simulations use Euler--Maruyama integration, while the PDE solutions are obtained via a finite element method. The implementation details and the source code are available in~\cite{votermodelcode}.
To compare the three formulations on equal footing, we require $\theta N$ to be an integer in the discrete model and replace $\theta$ by $\theta':=(\theta N+\tfrac{1}{2})/N$ in the SDE and PDE to account for the half-step offset; all stated thresholds refer to the original $\theta$.

\subsection{Single Parameter and Pairwise Comparisons}
\begin{figure}[t!]
    \centering
        \begin{center}
    \includegraphics[width=15cm]{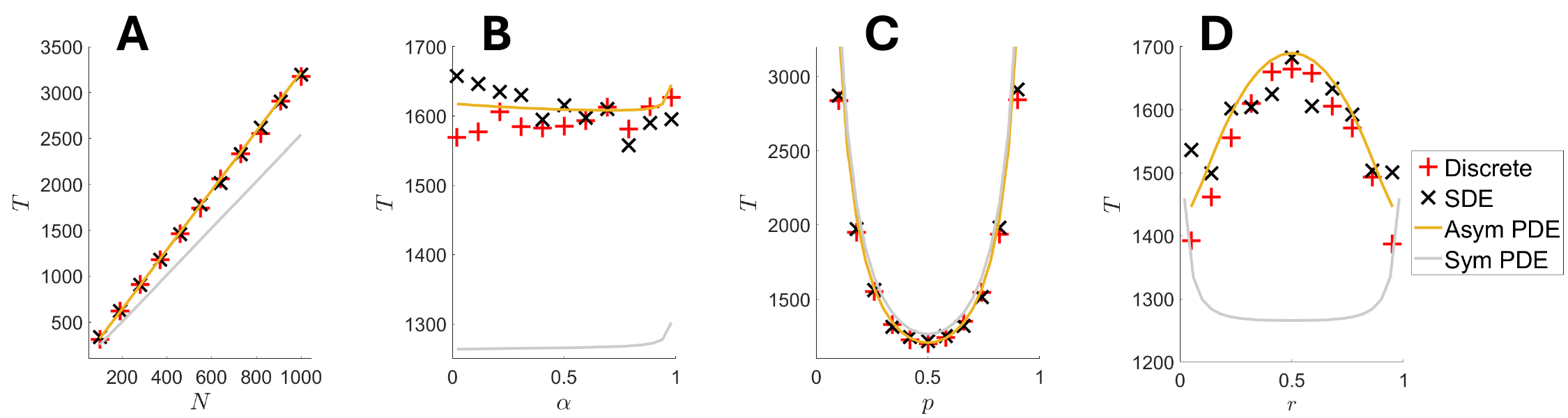} 
    \end{center}
    \caption{\textbf{Comparison between discrete, SDE, and PDE models across parameter sweeps.} Each panel overlays results from a symmetric baseline (shown in grey) and an asymmetric setting (shown in color): \textbf{(A)} total population size, $N$, \textbf{(B)}, modularity parameter, $\alpha$, \textbf{(C)}, flip probability, $p$, \textbf{(D)}, relative sizes, $r$. The symmetric case uses $y_1=y_2=p=\alpha=r=1/2$, while the asymmetric case uses $y_1=1/4$, $y_2=3/4$, $p=\alpha=3/4$, and $r=1/4$. In all panels, parameters not being varied are held fixed at these values, with $N=500$,  and $\theta=0.01$.}
    \label{fig:compsym3}
\end{figure}

Here we verify that the discrete, SDE, and PDE models produce consistent predictions by performing single-parameter experiments. Specifically, we examine how each of the four parameters $N$, $\alpha$, $p$, and the relative clique size $r := N_1 / N$ affects the system by varying them one at a time. We begin with a symmetric baseline where $p = \alpha = r = y_1 = y_2 = 1/2$, and then move to an asymmetric setting with $p = \alpha = y_2 = 3/4$ and $r = y_1 = 1/4$. Both cases are presented in Figure~\ref{fig:compsym3}. Across both settings, the three models are in good agreement, with deviations appearing only at extreme parameter values, where either the diffusion coefficients degenerate ($\alpha\to 1$, $p\to 0,1$) or finite-size effects limit the validity of the continuum approximation ($r\to 0,1$).
This confirms that the PDE analysis of the previous section captures the behavior of the discrete and SDE dynamics, and justifies our use of the PDE for the parameter explorations that follow.

We observe an approximately linear relation between $N$ and the mean consensus time $T$ (Figure~\ref{fig:compsym3}A). Asymmetry increases the slope: in Figure~\ref{fig:compsym3}A, the symmetric $T(N=1000) \approx 2500$, whereas asymmetric $T(N=1000) \approx 3250$.
By contrast, $\alpha$ has little effect on consensus time except near the disconnected limit $\alpha\to 1$, where $T$ increases sharply. In this regime, the cliques evolve nearly independently and may reach opposing consensus states; since heterogeneous interactions are absent, the system becomes trapped near the $(0,1)$ or $(1,0)$ states. Correspondingly, these points are singular for \eqref{eq:PDEforT}. In the symmetric case, $T$ increases in $\alpha$ but is nonmonotonic and convex in the asymmetric setting (Figure~\ref{fig:compsym3}B), indicating a stronger sensitivity to modularity away from symmetry. Asymmetry increases $T$ as well.
Figure \ref{fig:compsym3}C shows that the dependence on $p$ is symmetric about $p=1/2$, with divergence as $p\to 0$ or $p\to 1$. Thus, the primary driver of the consensus time is $p$: the farther $p$ is from $1/2$, the larger $T$ becomes. This is consistent with the asymptotic analysis, where the slow diffusive timescale is proportional to $[p(1-p)]^{-1}$. Symmetry or asymmetry does not seem to affect the time, as two PDE solutions almost coincide. 

Our final comparison examines how the relative sizes of the cliques affect the mean consensus time, as shown in Figure~\ref{fig:compsym3}D. Since the total population size is fixed, $r$ also reflects the ratio $N_1/N_2$. Once again, the three models agree well, with deviations only at extreme values of $r$, where the approximation breaks down for very small subpopulations (e.g., $N_1=10$).
The curves are approximately symmetric about $r = 1/2$, but the shape changes qualitatively between settings: in the symmetric case, $T$ is minimized at $ r= 1/2$, whereas in the asymmetric setting the minimum shifts toward $r$ closer to 0 or 1. In fact, in the asymmetric case, the maximum consensus time occurs at $r=1/2$. This reversal suggests that the role of equal clique sizes depends sensitively on the overall parameter regime. More broadly, the dependence of $T$ on $r$ interacts nontrivially with the other parameters, a point we return to in the next subsection.

\begin{figure}[t!]
    \centering
    \begin{center}
    \includegraphics[width=15cm]{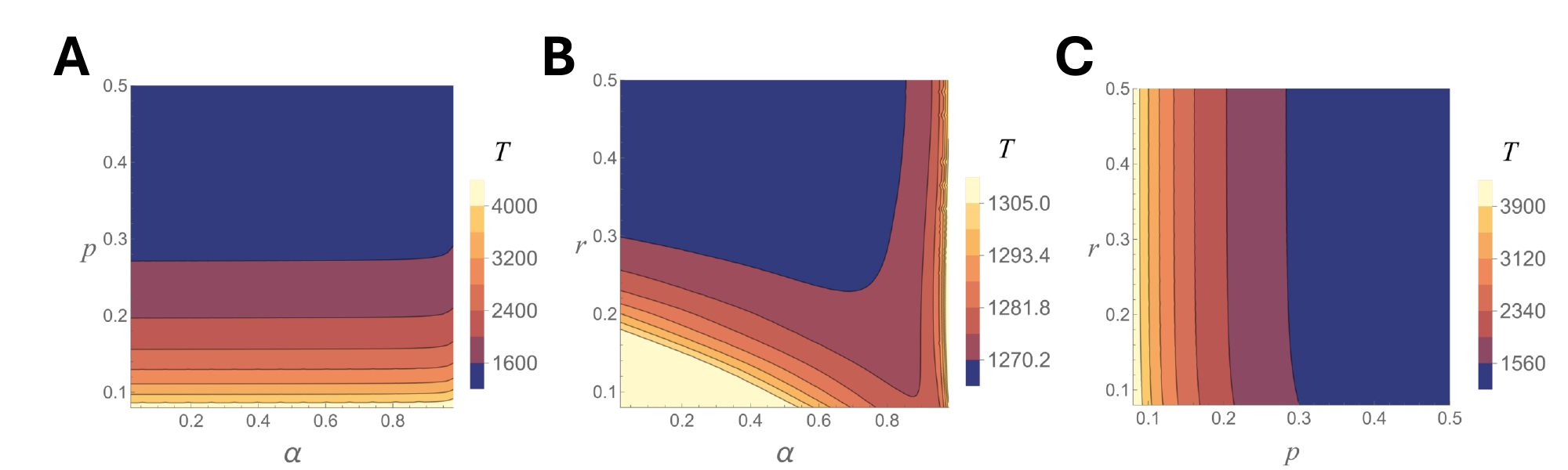} 
    \end{center}
    \caption{\textbf{Mean consensus time as a function of parameter pairs.} Contours of the PDE solution \eqref{eq:PDEforT} over \textbf{(A)}~$(\alpha,p)$ with $r=1/2$, \textbf{(B)}~$(\alpha,r)$ with $p=1/2$, and \textbf{(C)}~$(p,r)$ with $\alpha=1/2$. Remaining parameters $y_1=1/2$, $y_2=1/2$, $N=500$, and $\theta=0.01$.}
    \label{fig:paracombo}
\end{figure}

The single-parameter comparisons above suggest that $p$ acts largely as a uniform scaling factor, while $\alpha$ and $r$ interact in less obvious ways. To make this precise, we solve \eqref{eq:PDEforT} over each parameter pair $(\alpha,p)$, $(\alpha,r)$, and $(p,r)$, holding the third fixed along with $N=500$, $\theta=0.01$, and $y_1=y_2=1/2$ (Figure~\ref{fig:paracombo}). The contour plots confirm the scaling role of $p$: in Figures~\ref{fig:paracombo}A and~\ref{fig:paracombo}C, contours run nearly horizontal and vertical, respectively, so $p$ shifts $T$ uniformly without reshaping its dependence on $\alpha$ or $r$; slight deviations appear only near $\alpha\to 1$ and $r\to 0$. The more revealing panel is Figure~\ref{fig:paracombo}B: as $r$ decreases, $T$ becomes increasingly convex in $\alpha$, with a minimum at $r=1/2$ and a maximum in the corner of small $r$ and small $\alpha$---likely reflecting the low probability of selecting heterogeneous pairs in that regime. These results confirm that modularity matters most when clique sizes are highly unequal, motivating the detailed exploration of $\alpha$ in the next subsection.

The single-parameter results (Figure~\ref{fig:compsym3}) clearly indicate a hierarchy of influence. The simplest dependence is on $N$: consensus time grows linearly with population size, as one would expect from the diffusive timescale identified in the asymptotic analysis. The flip probability $p$ is the dominant control parameter---$T$ is symmetric about $p=1/2$, where it is minimized, and diverges toward both $p\to 0$ and $p\to 1$, in direct agreement with the slow-stage diffusion rate $p(1-p)$. Neither of these dependencies changes qualitatively between symmetric and asymmetric settings. The modularity parameter $\alpha$ tells a different story: in the symmetric case, $T$ is nearly flat across a wide range of $\alpha$ before diverging as the cliques decouple ($\alpha\to 1$), but once asymmetry is introduced the curve becomes convex, suggesting that an interior optimum may exist. The richest behavior belongs to the relative clique size $r$. In symmetric settings, equal-sized cliques ($r=1/2$) minimize consensus time; in asymmetric settings, this relationship reverses, and $r=1/2$ becomes a maximizer. This reversal, together with the emergent convexity in $\alpha$, points to a nontrivial interplay between modularity and population imbalance that we explore in detail in the next subsection.

\subsection{When Can Modularity Accelerate Consensus?}

The previous subsection established that modularity and population imbalance interact nontrivially, while $p$ acts as a uniform scaling factor. We now ask a sharper question: can a two-clique network reach consensus faster than a well-mixed population, and if so, is there an optimal level of inter-clique coupling? To expose these effects, we adopt polarized initial conditions ($y_1=1$, $y_2=0$), modeling groups that are initially sorted by opinion~\cite{trubowitz2005going, bail2018exposure,dandekar2013biased}, and rely on the PDE~\eqref{eq:PDEforT} throughout.
As a baseline, we compare against a well-mixed model on the complete graph with the same update rules. The corresponding mean consensus time admits an explicit solution~\cite{redner2019reality,ewens2004mathematical}:
\begin{equation}\label{eq:T1Dsol}
    T_{1D}:=-\frac{N}{p(1-p)}\big[x_0\log(x_0)+(1-x_0)\log(1-x_0)-\theta\log(\theta)-(1-\theta)\log(1-\theta)\big].
\end{equation}
where $x_0 := (y_1 N_1 + y_2 N_2)/N$ is the initial fraction of opinion-A holders.

\begin{figure}[t!]
\centering
\begin{center} \includegraphics[width=15cm]{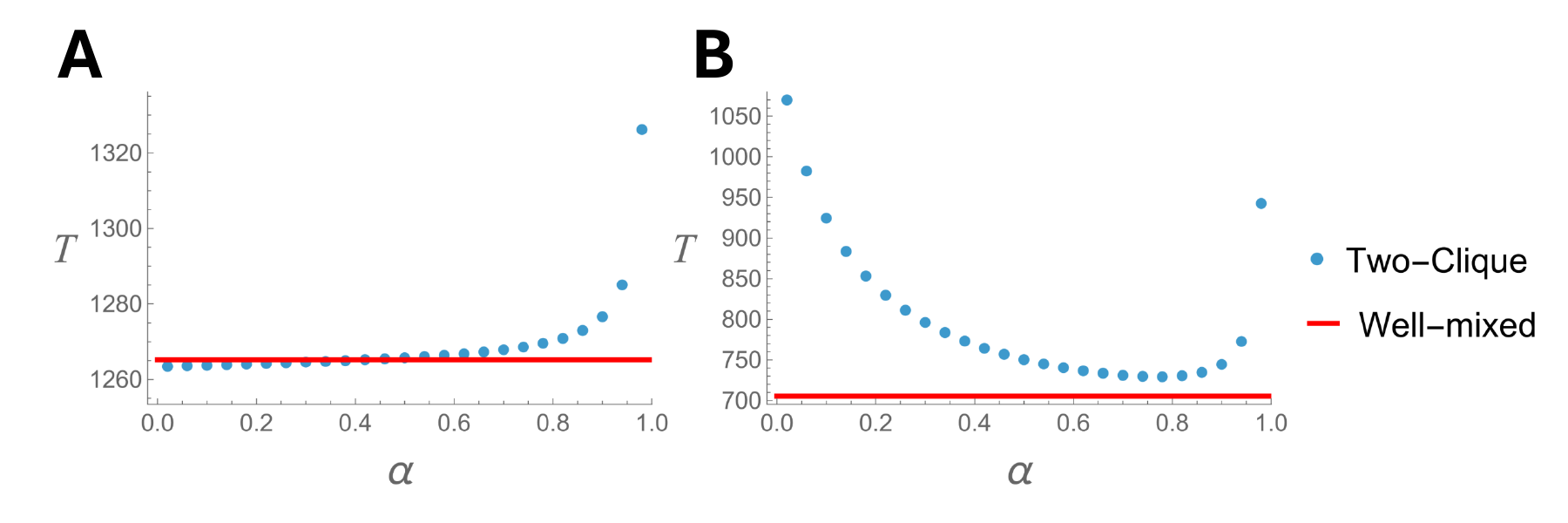} \end{center}
    \caption{\textbf{Mean consensus time: two-clique versus well-mixed model}.
    \textbf{(A)}~Symmetric case, $N_1=N_2=250$, $p=1/2$.
    \textbf{(B)}~Asymmetric case: $N_1=50$, $N_2=450$, $p=1/4$.
    Remaining parameters: $\theta=0.01$, $y_1=1$, $y_2=0$.
    The red line is the well-mixed solution~\eqref{eq:T1Dsol}; blue dots are the two-clique PDE solution~\eqref{eq:PDEforT}.}
    \label{fig:compare}
\end{figure}

In the symmetric case (Figure \ref{fig:compare}A), the two-clique model is marginally faster at small $\alpha$ where inter-clique edges preferentially select heterogeneous pairs under polarized initial conditions; the advantage is confined to the fast alignment stage and amounts to less than 1\% of the total consensus time.
The asymmetric case (Figure~\ref{fig:compare}B) is more revealing: $T(\alpha)$ becomes convex with a clear interior minimum, indicating an optimal level of inter-clique coupling exists.

The U-shape can be understood through a small-$r$ scaling argument. Let $r := N_1/N \ll 1$. In this limit, the edge fractions simplify to
\begin{equation*}
\gamma_1 = \mathcal{O}(r^2), \qquad \gamma_2 \approx 1 - \frac{2(1-\alpha)r}{\alpha}, \qquad \gamma_3 \approx \frac{2(1-\alpha)r}{\alpha}.
\end{equation*}
Since $\gamma_3 = \mathcal{O}(r)$, inter-clique interactions become rare. Substituting into the drift expressions from Section~2.2 and rescaling by $N/2$ gives
\begin{equation*}
\mu_1 \approx \frac{(1-\alpha)p}{\alpha}(x_2 - x_1), \qquad 
\mu_2 \approx \frac{(1-\alpha)p\,r}{\alpha}(x_1 - x_2).
\end{equation*}
The small clique tracks the large one at rate $(1-\alpha)p/\alpha$, while the large clique is slower by a factor of $r$. The corresponding  alignment timescale is the reciprocal of this rate,
\begin{equation*}
\tau_{\mathrm{fast}} \sim \frac{\alpha}{(1-\alpha)p},
\end{equation*}
which favors small $\alpha$. The leading-order diffusion coefficient for $x_1$ scales as
\begin{equation*}
D_{11} \sim \frac{(1-\alpha)\,p}{\alpha\,r\,N}\big(x_1(1-x_2)+x_2(1-x_1)\big),
\end{equation*}
which diverges as $\alpha \to 0$: reducing modularity amplifies 
fluctuations in the small clique. On the alignment manifold 
($x_1\approx x_2\approx x$), the parenthetical factor reduces to 
$2x(1-x)$, recovering the expression used below.

After the fast stage, the system evolves along the alignment manifold $x_1 \approx x_2 =: x$ with effective initial fraction $x_0 \approx r$ (under polarized initial conditions $y_1 = 1$, $y_2 = 0$). At leading order, the slow-stage dynamics reduce to the well-mixed model on the large clique, giving a total consensus time of approximately
\begin{equation*}
T \approx \underbrace{\frac{\alpha}{(1-\alpha)p}}_{\text{fast stage}} \underbrace{- \frac{N}{p(1-p)}\big[r\log r + (1-r)\log(1-r) - \theta\log\theta - (1-\theta)\log(1-\theta)\big]}_{\text{slow stage}}.
\end{equation*}
The first term decreases with $\alpha$; the second is independent of $\alpha$ and typically dominates. However, a correction to the slow stage restores $\alpha$-dependence. On the alignment manifold, $x_1+x_2-2x_1x_2\approx 2x(1-x)$ and $x_i(1-x_i)\approx x(1-x)$, so the small-$r$ diffusion components from Section~\ref{sec:cont} reduce to
\begin{align*}
    D_{11} &= \frac{(1-\alpha)\,p\,x(1-x)}{\alpha\,r\,N}, \qquad
    D_{22} \approx \frac{p(1-p)\,x(1-x)}{N}, \qquad
    D_{12} = -\frac{(1-\alpha)\,p^2\,x(1-x)}{\alpha\,N}.
\end{align*}
Since $x = rx_1+(1-r)x_2$ is a linear combination, $D_{\mathrm{eff}} = r^2D_{11}+2r(1-r)D_{12}+(1-r)^2D_{22}$. The dominant contribution is $(1-r)^2D_{22}\approx p(1-p)\,x(1-x)/N$, the standard well-mixed diffusion. The term $r^2D_{11} = r(1-\alpha)\,p\,x(1-x)/(\alpha N)$ provides an $\mathcal{O}(r)$ correction (the $r^2$ cancels one power of $r$ in $D_{11}$), while $2r(1-r)D_{12}$ is also $\mathcal{O}(r)$ but carries $p^2$ rather than $p(1-p)$ and is dropped at leading order. This gives
\begin{equation}\label{eq:Deff}
D_{\mathrm{eff}} \approx \frac{p\,x(1-x)}{N}\left[(1-p) + \frac{r(1-\alpha)}{\alpha}\right].
\end{equation}
Replacing the well-mixed diffusion in the slow stage with~\eqref{eq:Deff} yields the corrected estimate
\begin{equation}\label{eq:Tsmallr2}
T \approx \frac{\alpha}{(1-\alpha)p} - \frac{N}{p\!\left[(1-p) + \frac{r(1-\alpha)}{\alpha}\right]}\big[r\log r + (1-r)\log(1-r) - \theta\log\theta - (1-\theta)\log(1-\theta)\big].
\end{equation}
Equation~\eqref{eq:Tsmallr2} makes the competing roles of $\alpha$ explicit. As $\alpha$ decreases from $1$, the fast-stage cost drops and the enhanced diffusion~\eqref{eq:Deff} shortens the slow stage, so $T$ decreases, consistent with the right branch of the U observed in simulations. For large $\alpha$ the correction term $r(1-\alpha)/\alpha$ vanishes, the slow stage reverts to the well-mixed value, and the fast-stage cost diverges as $\alpha\to 1$. The observed optima ($\alpha^*\approx 0.7$--$0.9$) lie in the transitional regime where both contributions are comparable.

\begin{figure}[t!]
    \centering
        \includegraphics[width=0.4\linewidth]{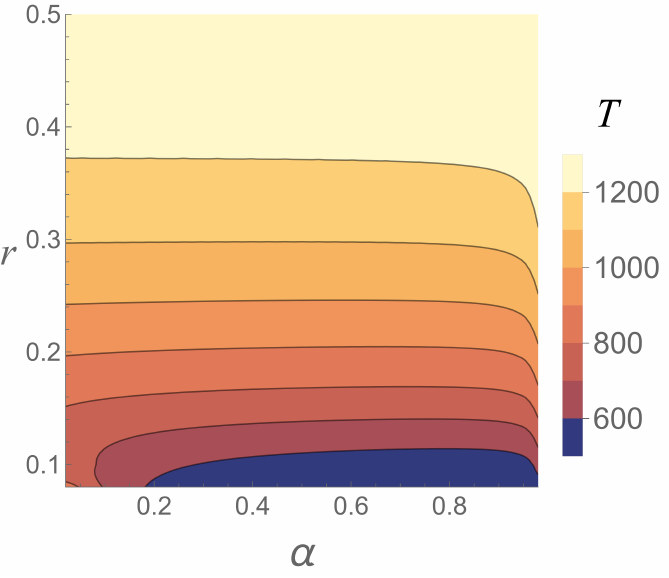}
    \caption{\textbf{Mean consensus time as a function of $\alpha$ and $r$
under polarized initial conditions.} Contours of the PDE solution~\eqref{eq:PDEforT} with $y_1=1$, $y_2=0$, $N=500$, $p=1/2$, $\theta=0.01$. Compare with Figure~\ref{fig:paracombo}B, which uses balanced initial conditions.}
    \label{fig:rvsalpha}
\end{figure}

Figure~\ref{fig:rvsalpha} confirms this picture under polarized initial conditions ($y_1=1$, $y_2=0$): as $r$ decreases, $T(\alpha)$ becomes increasingly convex, with both the depth of the minimum and the overall range of $T$ growing. Compared with the balanced initial conditions of Figure~\ref{fig:paracombo}B, polarization shifts the minimum to smaller $\alpha$ and amplifies the penalty for near-disconnection. As in the previous subsection, varying $p$ shifts $T$ uniformly without altering this structure.

We next examine how initial polarity affects the optimal $\alpha$ and the corresponding $T$. Fixing $r=1/10$ (Figure \ref{fig:rvsalpha}), both the optimal $\alpha$ and the corresponding 
minimal $T$ decrease as $|y_1 - y_2|$ increases. The small-$r$ scaling argument above provides intuition: stronger polarization means the system starts further from alignment, increasing the relative contribution of the fast stage to the consensus time. To compensate, the optimum shifts toward smaller $\alpha$, which accelerates drift at the cost of larger fluctuations in the small clique. The decrease in minimal $T$ is more straightforward: as the minority shrinks, fewer agents must be converted to obtain consensus.

\begin{figure}[t!]
\centering
\begin{center} \includegraphics[width=15cm]{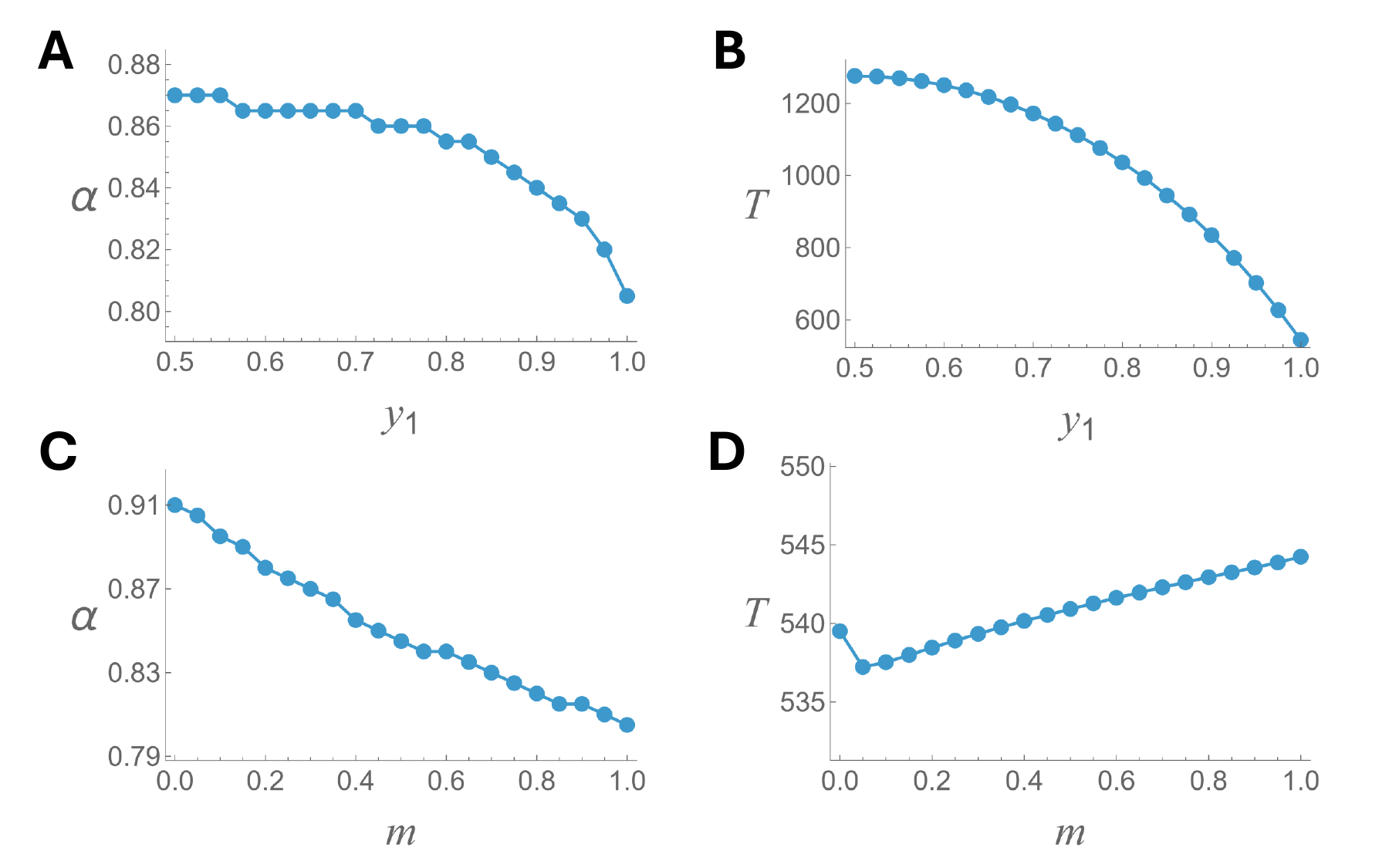} \end{center}
    \caption{\textbf{Effect of initial polarization on optimal coupling and consensus time.} \textbf{(A,\,B)}~Unconstrained polarization: $y_2=1-y_1$ varies freely. \textbf{(C,\,D)}~Constrained polarization: the total number of minority-opinion holders is fixed at $M=50$ and redistributed between cliques via $m:=y_1M/N_1$. In all panels, $N_1=50$, $N_2=450$, $p=1/2$, $\theta=0.01$. As clique sizes become more comparable, the constrained-polarity trends weaken and eventually reverse.}
    \label{fig:optim_combined}
\end{figure}

A natural follow-up question is whether this acceleration persists when the total number of minority-opinion holders is conserved, rather than allowing initial fractions to vary freely. A more realistic scenario fixes the total number of minority-opinion holders $M$ and asks how they should be distributed: concentrated in the small clique, spread across both, or somewhere in between? To model this, we define the \textit{constrained polarity} parameter $m:=y_1N_1/M$, so that $y_1=mM/N_1$ and $y_2=(1-m)M/N_2$ (Figure~\ref{fig:optim_combined}C,D).

Under this constraint, the optimal $\alpha$ increases roughly linearly with $m$, and consensus generally takes longer than in the unconstrained case. A clear minimum in $T$ emerges at an intermediate value $m\approx 0.05$, which can be understood through the competing timescales identified in Section~3: small $m$ concentrates the minority in the small clique, maximizing the opinion gap $|y_1-y_2|$ and thereby shortening the fast alignment stage ($\tau_{\mathrm{fast}}\sim\alpha/[(1-\alpha)p]$), while simultaneously pushing the post-alignment fraction $x_0\approx M/N$ toward zero, where the slow diffusion to consensus is fastest. Below this optimum, fluctuation costs in the small clique ($D_{11}\sim(1-\alpha)/(\alpha\,r\,N)$) dominate and consensus slows; above it, the minority is diluted across both cliques, weakening the alignment advantage without compensating gains in the slow stage. In practical terms, the minimum suggests that concentrating most, but not all, of the minority in the smaller clique best balances rapid alignment against efficient diffusion toward consensus. As clique sizes become more comparable, these trends weaken and eventually reverse, with the optimal $\alpha$ dropping and the minimum in $T$ shifting toward $m\approx 1$. An explicit expression for the optimal $\alpha$ under constrained polarity remains an open problem. The small-$r$ scaling correctly predicts the qualitative trends, but the observed optima ($\alpha^*\approx 0.7$--$0.9$) lie closer to the disconnection limit than the leading-order balance suggests, indicating that the slow-stage cost and the $\alpha\to 1$ divergence play a significant role in setting the precise optimum.

To summarize: modularity has little effect on consensus time in symmetric settings, where the well-mixed model is nearly optimal. Its influence becomes significant under population imbalance and polarized initial conditions, where an optimal level of inter-clique coupling emerges from the competition between fast alignment drift and small-clique fluctuations.

\section{Discussion}
In this paper, we studied a modular voter model on two cliques, analyzing how modularity, population imbalance, and initial polarization shape the time to consensus. We derived SDE and Fokker-Planck approximations of the discrete process and confirmed their agreement numerically across a range of parameter regimes.
The multiscale asymptotic analysis of the Fokker-Planck equation revealed two distinct stages: a fast alignment phase in which inter-clique coupling drives exponential synchronization of opinion fractions at rate $2p(1-\alpha)$, followed by a slow diffusion along the alignment manifold governed by $p(1-p)$ and independent of $\alpha$. For unequal clique sizes, a complementary small-$r$ scaling analysis identified the competing mechanisms that give rise to an optimal modularity: reducing $\alpha$ accelerates alignment ($\tau_{\mathrm{fast}} \sim \alpha/[(1-\alpha)p]$) but amplifies fluctuations in the small clique ($D_{11} \sim (1-\alpha)/(\alpha\,r\,N)$), producing the U-shaped dependence of consensus time on $\alpha$ observed in simulations. The corrected decomposition~\eqref{eq:Tsmallr2} captures the qualitative structure of the U-shape, with the observed optima ($\alpha^*\approx 0.7$--$0.9$) lying in the transitional regime where fast-stage and slow-stage contributions are comparable.

Numerical exploration confirmed that $p$ acts primarily as a uniform scaling factor, while the interaction between $\alpha$ and the relative clique size $r$ produces the richest behavior. Under polarized initial conditions, an optimal $\alpha$ emerges that decreases with increasing polarization. Under constrained polarity, where a fixed minority is redistributed across cliques, an interior optimum in the distribution parameter $m$ reflects the balance between fast alignment and slow diffusion. The observed optima ($\alpha^* \approx 0.7$--$0.9$) lie closer to the disconnection limit than leading-order scaling predicts, indicating that higher-order corrections are needed for quantitative agreement.

Several directions remain open. The small-$r$ scaling could be extended to obtain sharper bounds on the optimal $\alpha$, potentially through matched asymptotic expansions that connect the fast and slow regimes. Well-posedness of the mean consensus time PDE~\eqref{eq:PDEforT} near the singular points $(0,1)$ and $(1,0)$ deserves rigorous treatment. The tolerance parameter $\theta$, which we have not analyzed in detail, may itself influence the optimal coupling in ways worth exploring. Extensions to more than two cliques, non-binary opinions, and heterogeneous flip probabilities would bring the model closer to realistic settings. Most importantly, calibrating the model against empirical data on opinion dynamics or behavioral spread in structured populations would test whether the mechanisms identified here operate in practice.
More broadly, the drift-noise competition identified here for two cliques may generalize to hierarchical or multilayer networks, where similar timescale separations arise between local and global consensus~\cite{masuda2014voter}, and the introduction of zealots or committed minorities could further enrich the dynamics~\cite{mobilia2003does}. \\

\noindent
{\bf Acknowledgement:}  This work was partially funded by NSF-DMS-2042413, AFOSR MURI FA9550-22-1-0380, NSF-DMS-2207700, and NSF-DMS-2527338.

\bibliographystyle{unsrt}
\bibliography{ref}

\end{document}